# Graphical Abstract

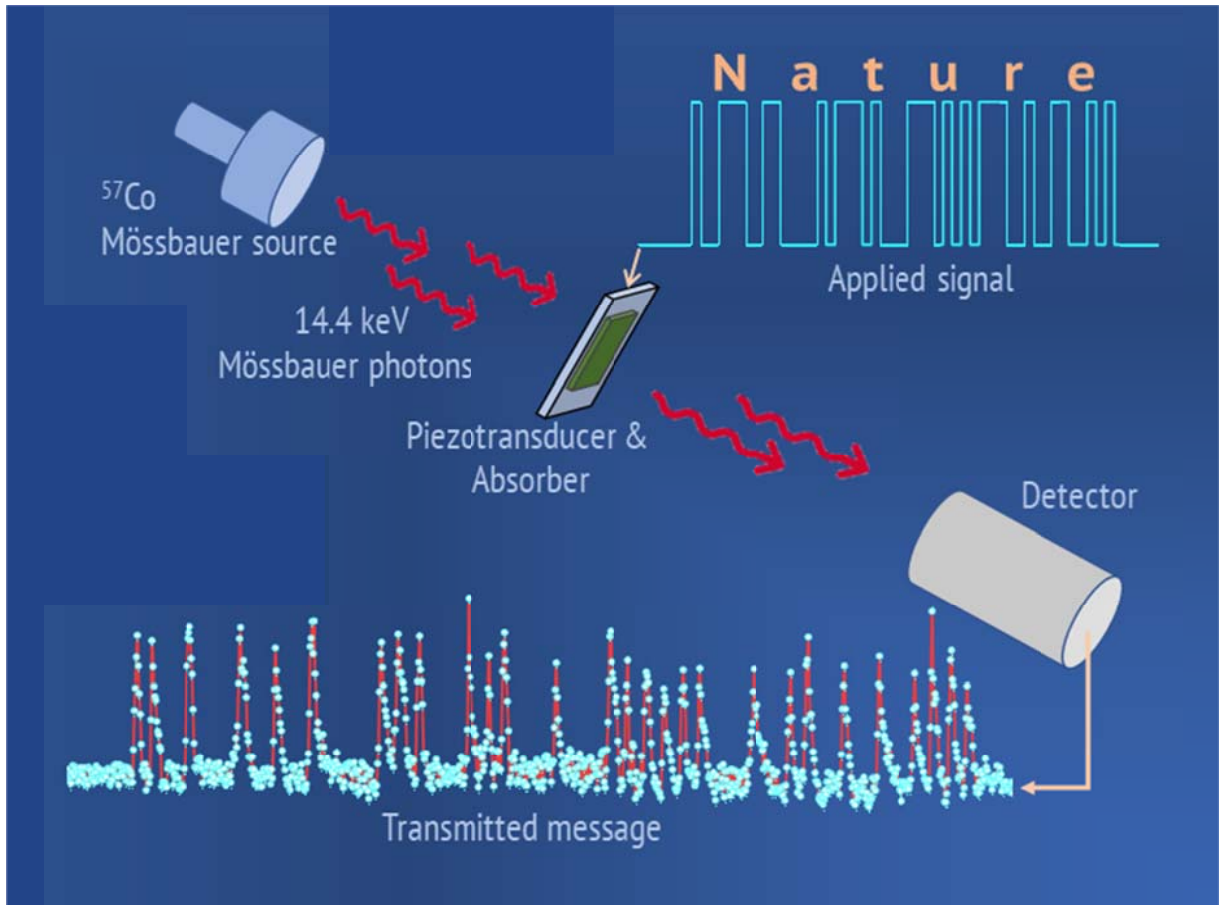

# Gamma-protocol for secure transmission of information

Rustem Shakhmuratov[*], Almaz Zinnatullin, and Farit Vagizov

Kazan Federal University, Kazan 420008, Russia

**Secure communication that allows only the sender and intended recipient of a message to view its content has a long history. Quantum objects, such as single photons are ideal carriers for secure information transmission because, according to the no-cloning theorem[1], it is impossible to create an identical and independent copy of an arbitrary quantum state while its detection leads to the information distortion. BB84[2,3] is the first quantum cryptography protocol for a quantum key generation and distribution, based on single photon sources. This quantum key is used for coding and decoding of classical information. We propose completely different protocol based on a stochastic decay of an ensemble of radioactive nuclei randomly emitting a stream of gamma-photons. We experimentally demonstrate a method how to transmit classical information containing binary bits (0 or 1) with the help of this stream. Transmission is organized such that eavesdropping is impossible since the presence of information in the stream of randomly emitted gamma-photons can be hidden. Reading of this information needs precise knowledge of the repetition rate of its sending in advance. It is unrealistic for the eavesdropper to disclose this rate, and without knowledge of this parameter it is impossible to make the transmitted information visible.**

[*]Corresponding author: shakhmuratov.rustem@gmail.com

Quantum gamma-optics is rapidly developing field of science. Gamma-photons with energies ranging from 10 to 100 keV have many important advantages for applications. The term gamma-ray radiation is commonly used to refer to photons emitted by atomic nucleus, while X-ray radiation in the same frequency band is emitted by electrons. Many interesting experiments were implemented with narrow linewidth gamma-photons. Among them are coherent control of the waveforms of recoilless γ-ray photons[4,5], electromagnetically induced transparency in a cavity[6], vacuum-assisted generation of atomic coherences[7], collective Lamb shift[8], parametric down-conversion in the Langevin regime[9], subluminal propagation of X-rays[10] and gamma-rays[11], single-photon revival in nuclear absorbing multilayer structures[12], gamma-echo[13,14], high-resolution spectroscopy beyond homogeneous linewidth[15,16], and sub-angstrom spatial resolution topography of mechanical displacements in the vibrating thin-solid films[17-19] (see a review[20] for other related works).

Traditional quantum optics deals with optical photons, which are widely used in the new, rapidly developing fields of quantum cryptography[21] and quantum communication[22]. Quantum gamma-optics have important potential advantage over quantum optics since detectors for γ-photons have high efficiency due to their high energy and low noise because of vanishingly small number of false counts. The most popular nucleus in γ-domain, commonly used in solid-state spectroscopy, is $^{57}$Fe with 14.4-keV Mössbauer transition for γ-photons with the wavelength of 86 pm. The transition quality factor $Q$, which is the ratio of the resonance frequency to the linewidth, is very high for these nuclei. Indeed, the linewidth is 1.1 MHz for $^{57}$Fe, which gives $Q = 3 \times 10^{12}$. High efficiency detectors and narrow resonant linewidth for 14.4-keV photons enable implementation of many interesting experiments in quantum γ-optics.

Another very important advantage is the possibility to prepare sources of single γ-photons emitted in time intervals well exceeding the coherence length of a single-photon wave-packet. First experiment on γ-echo demonstrating a striking effect of a single photon reviaval[13] was implemented with a source containing radioactive $^{57}$Co nuclei incorporated into Rh foil. The source activity was 50 kBq corresponding to the emission of $5 \times 10^4$ photons per second. The decay scheme of radioactive $^{57}$Co is shown in Fig. 1. This nuclide has half-life of 272 days and decays by electron capture to the second excited state of $^{57}$Fe, which mainly emits in cascade a 122 keV precursor photon and a 14.4 keV Mössbauer (narrow linewidth) photon via the first excited state of $^{57}$Fe. The lifetime of this state $|1\rangle$ is 141 ns. Within this time, mean number of emitted Mössbauer photons is $7 \times 10^{-3}$ ensuring that emission of two 14.4 keV photons or more is negligibly small.

We propose to transmit classical information using a stream of randomly emitted single photons and develop a secure information-transmission method that prevents reading by an eavesdropper. Our protocol is based on generation of a pulsed signal in a temporal waveform of a single photon wave-packet. This method is experimentally demonstrated using Mössbauer γ-photons emitted by decaying $^{57}$Fe. Below we explain the basic principles of our method.

In time-domain Mössbauer spectroscopy, a 14.4 keV γ-photon wave-packet, emitted by the $^{57}$Fe excited state is reconstructed as an exponentially decaying radiation field with a sharp rising leading edge by use of time-delayed coincidence counts of two detectors. Detection of the122 keV precursor photon starts the clock defining time when a 14.4 keV excited state nucleus is formed in the source. Detection of a 14.4 keV Mössbauer photon stops the clock. Time distribution of the intervals between these counts is obtained using a data analyzer (DA), which consists of time to amplitude converter (TAC) working in start-stop mode and data acquisition module. When a photon with an energy of 14.4 keV is transmitted through a thick resonant absorber containing $^{57}$Fe nuclei, the wave packet of Mössbauer photon decays considerably faster than the source radiation due to its multiple scattering by resonant nuclei[23]. Stepwise displacement of the emitting source[13] or the resonant absorber [14] by a half wavelength induces coherent flash of Mössbauer radiation, named the γ-echo[13]. The amplitude of this echo can be much higher than the amplitude of the source radiation.

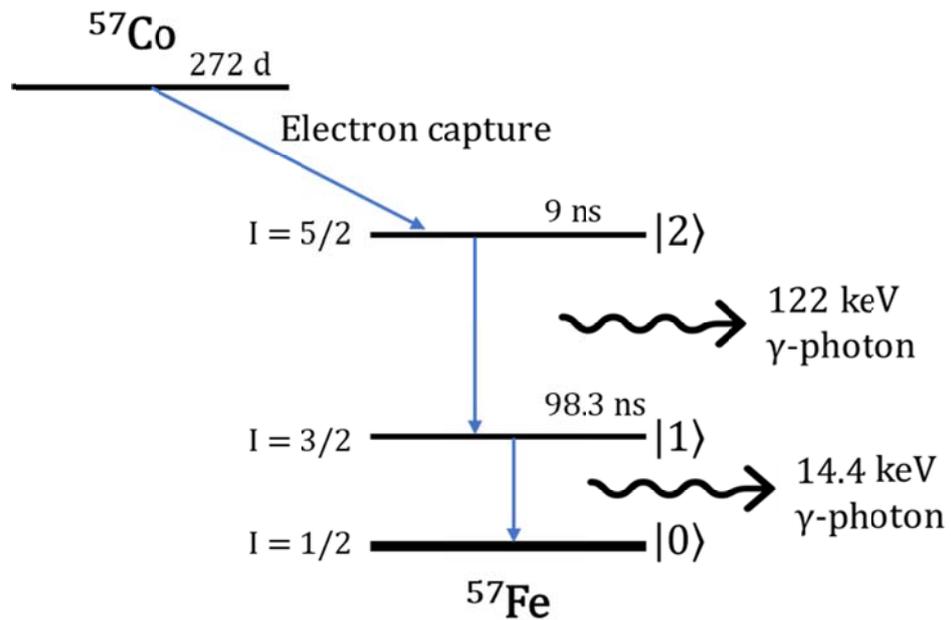

Fig.1 Decay scheme of the $^{57}$Co nuclide. It converts to $^{57}$Fe in the second excited state by electron capture. Half-life (lifetime multiplied by ln2) is indicated for each state in the scheme. The less probable direct 136.5 keV transition from $|2\rangle$ to $|0\rangle$ is not shown.

This effect is explained by the following arguments. The source radiation propagating in the resonant absorber experiences coherent scattering by resonant particles in forward direction. Coherently scattered radiation is opposite in phase with the incident field[13,24]. Therefore, these radiation fields interfere destructively resulting in attenuation of the propagating field. If, at any moment of time when coherently scattered field is sufficiently developed to cause the field attenuation at the exit of the absorber, the phase of the incident field is rapidly changed by $\pi$, then the incident and coherently scattered fields interfere constructively producing a coherent flash. This coherent flash was also observed in optical domain for coherent continuous-wave radiation field[25].

When the transmitted γ-radiation is measured as a function of time synchronized with the phase of the source [26,27] or absorber[28] motion, short, enhanced pulses of γ-radiation are also clearly observed. In this case, the precursor photon is not detected. In experiments reported in refs. 26,27 and 28, piezoelectric transducers are used displacing the source or absorber. Rectangular shaped voltage with sharp leading and trailing edges is applied to the transducer. These pulses produce a half wavelength displacement causing desirable phase shift of the radiation field seen by the absorber nuclei. In the experiments[26-28], rising and falling edges of the rectangular voltage produce γ-radiation pulses. It is remarkable that a stream of randomly-emitted resonant γ-photons produces pulses at well-defined time intervals. This is because photons, which are emitted at time when the source or absorber do not experience displacement, are simply attenuated by the absorber, while photon wave-packets experiencing absorber displacement during their propagation are compressed into pulses.

We propose to apply this technique for transmission of classical information and experimentally implement the method, which we name γ-protocol. We demonstrate this protocol with single-photon-radiation source, which is $^{57}$Co incorporated into rhodium matrix. As an absorber we use a 10-μm-thick stainless steel-foil with a natural abundance (~2%) of $^{57}$Fe. This foil is glued on the polyvinylidene fluoride piezotransducer producing desired displacements of the absorber.

As an example text for transmission, we choose word "Nature" in the sense of the phenomena of the physical world in the totally, which humans study and use for their own needs. We employed extended ASCII codes for the text, which give binary bits 01001110 01100001 01110100 01110101 01110010 01100101. They are coded by a train of rectangular voltage pulses of varying duration. The total length of the pulse series is $T_{info} = N\tau$, where $N$ is the total number of bits and time scale is divided into time bins $\tau = 347$ ns. We prescribe bit 1 for time intervals where voltage pulse is present, and 0 for those where the voltage is zero, see Fig. 2(a).

Fig. 3 shows a schematic arrangement of the source, absorber, detector, and electronics used in our experiment. Binary sequence coding our text is observed as a train of pulses shown in Fig. 2(b). From this train it is easy to reconstruct the corresponding train of rectangular pulses of the applied voltage and then read the message.

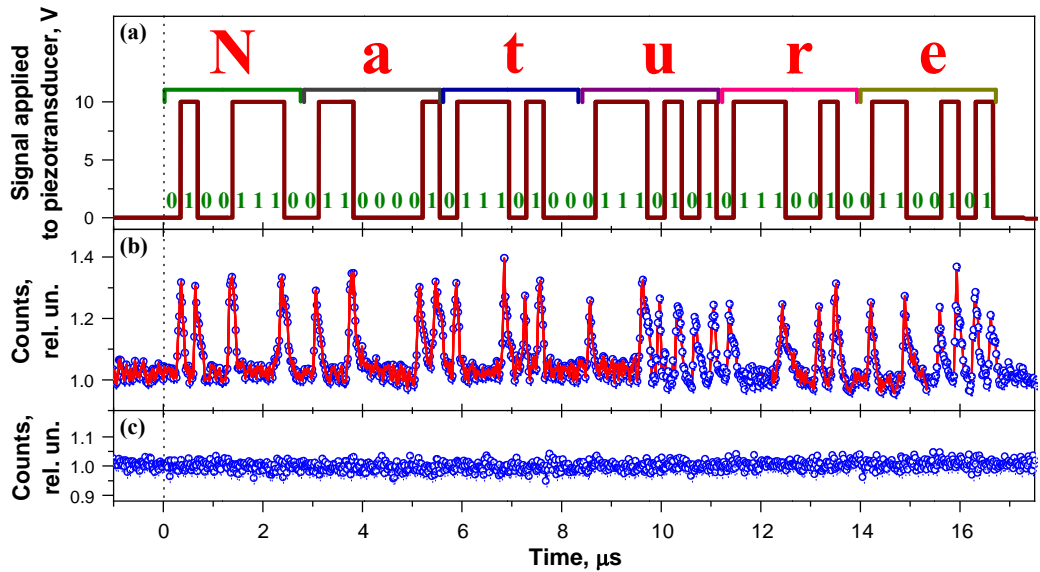

Fig. 2 (a) A train of rectangular voltage pulses coding binary bits 01001110 01100001 01110100 01110101 01110010 01100101. Signal transmission time is divided into time-slots equal to the number of bits in the message. Bits 1 and 0 correspond to the time-slots in which the voltage is turned on and off, respectively. (b) The sequence of γ-pulses generated by the absorber displacements produced by the sequence of the rectangular voltage pulses. Each pulse is induced by the leading and trailing edges of the voltage pulses. From this series of γ-pulses it is possible to unambiguously reconstruct a series of voltage pulses and message bits. (c) Time-domain spectrum for the case when the repetition frequencies of information sequence and Start signals initializing TAC are different by 0.01 Hz.

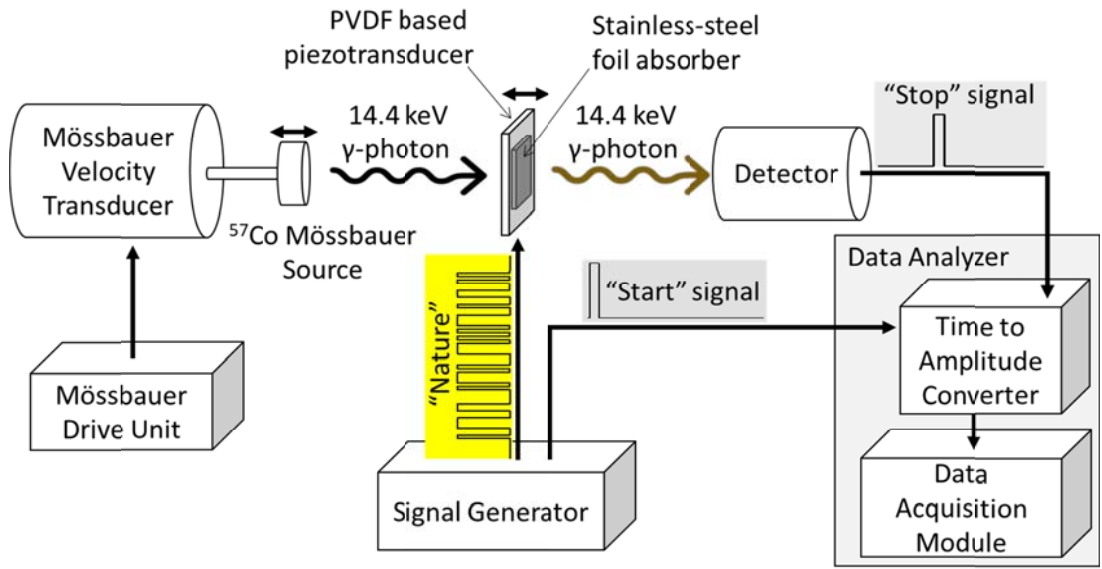

Fig. 3. Simplified scheme of the experimental setup for γ-protocol. $^{57}$Co Mössbauer source emits 14.4 keV photons and velocity transducer controls tuning the source in resonance with the stainless-steel foil absorber containing $^{57}$Fe nuclei. Signal generator produces a train of rectangular voltage pulses. Detector sends "Stop" signals to a time to amplitude convertor, which is synchronized with the signal generator by "Start" signal. Data acquisition module reproduces the generated γ-pulses.

The train of $\gamma$-pulses is obtained by collecting many counts of $\gamma$-photons, which are transmitted through the absorber experiencing fast displacements. They are produced by a sequence of rectangular-voltage information-pulses, which are coupled to "Start" signals. This sequence is repeated many times. Our TAC is initialized also by a periodic sequence of "Start" signals with the same repetition rate. Data acquisition module distributes counts over channels, each of which is responsible for a specific time slot. The repetition rate $\Omega_S = 1/T_S$ of the train of the information pulses must precisely match the repetition rate of "Start" signals initializing TAC. Since overall data acquisition time is much longer than $T_S$, any mismatch of these frequencies will cause random, almost homogeneous data spreading across channels, and then data will be lost. Example of this mismatch is shown in Fig.2(c), where the signal repetition rates is $\Omega_S = 50$ kHz while the repetition rate of signals initializing TAC is different from $\Omega_S$ by 0.01 Hz. This is the basic point of our $\gamma$-protocol. Reading of the message is impossible for undesirable recipient if eavesdropper does not know in advance this frequency with extremely high precision.

The presence of information in the $\gamma$-photon stream can be hidden. First, time when the information will be send and its repetition rate can be specified in advance by the sender and the intended recipient. Second, the information transmission by absorber displacements will cause a small change of the baseline of the photon counts due to slight bleaching of the absorber induced by displacements (see details in Supplementary Information). To hide this effect the sender can place a proper filter just before the information sending and remove it at the end. Third, it is not necessary to synchronize precisely the beginning of the information sending with information receiving. For example, if we keep exactly the same repetition rate of information coding and decoding but introduce a constant time shift between writing and reading of this information, we can still read true message. To correctly decode the information, a message may may begin with a code signal created by a rectangular pulse of a duration shorter than $\tau$ two or more times. This pulse can be easily found by the intended recipient in the train of $\gamma$-pulses. Rearranging the message such that this pulse will be placed first and the following information after, one can reconstruct the message properly. The end of the message can be marked by the doubled code signal. Otherwise, one can use "Start" (STX) and "End" (EXT) bytes from ASCII codes to indicate the beginning and the end of the message.

Transmission distance of $\gamma$-photons is limited despite their high penetration even through materials that are opaque for optical photons. This is because $\gamma$-photons are emitted by nuclei into a $4\pi$ solid angle and radiation intensity decreases inversely proportional to the distance squared. Meanwhile, the famous $\gamma$-photon experiment measuring the gravitational redshift[28] where carried out at a distance of 22.5 m using a $^{57}$Co radiation source with activity 1.48 x 10$^{10}$ Bq. This distance can be essentially increased with the reflecting and focusing elements rapidly developing in the soft $\gamma$- and X-ray domain[30,31].

The proposed method can be extended to the optical domain using luminescent sources emitting according to the stochastic Poisson process. Coupling of this field to the optical fiber allows significant increasing of the transmission distance. Moreover, the bitrate of the proposed coding method of optical photons can be essentially increased by modification[32], which employs $2\pi$ phase-shift of the field by electro-optical modulator instead of $\pi$ phase-shift used in $\gamma$-echo.

METHODS SUMMARY

The method of secure information transmission using single photons is based on analytical solution of Maxwell-Bloch equations (MBE) for propagation of an exponentially decaying wave-packet with sharp rising leading edge through resonantly absorbing medium. The phase of this wave packet is abruptly changed by $\pi$ at particular times. This is taken into account in the MBE-solution derived in the rotating wave and linear response approximations. The solution is obtained with the help of the Green's function derived in refs. 23,33 for classical field and in ref. 34 for quantum field. Then, this solution is averaged over time of photon emission by the source nuclei. The result is analyzed numerically. The shortest appropriate time interval between phase shifts is found. The influence of the optical thickness of the absorber on the induced signal is analyzed.

We use experimental apparatus conventional for time-domain Mössbauer spectroscopy. We used commercial arbitrary waveform generator producing a sequence of rectangular-shape RF pulses of designed duration and spacing for coding the transmitted information.

**Acknowledgements** This work was supported by RSF (grant no. 23-22-00261).

**Author Contributions** F.V. suggested the idea, supervised the project, developed the experimental methods, designed the experiment and derived all the experimental results. A.Z. performed the experiments, worked with data and their presentation. R.S. invented the concept of *γ*-protocol and wrote the paper. All authors discussed the results and edited the manuscript.

**Author Information** Correspondence and requests for materials should be addressed to R.S. (shakhmuratov.rustem@gmail.com)

**Supplementary Information**

Here we give a brief overview of the theoretical basis of the physical processes used in γ-protocol. The propagation of Mössbauer γ-photon wave packet through a resonant absorber can be described semi-classically[23,33] or quantum mechanically[34]. This wave packet is treated as a damped electromagnetic field

$$a_0(t - t_0) = \theta(t - t_0)e^{-(i\omega_S + \gamma)(t - t_0)}, \qquad (1)$$

where $a_0(t - t_0)$ can be considered as a conditional probability amplitude of the second 14.4 keV photon in a cascade decay of $^{57}$Fe nucleus, or amplitude of classical field, $\omega_S$ is the frequency of the 14.4 keV photon, $2\gamma$ is the decay rate of the 14.4 keV state, $t_0$ is time when this state is formed, and $\theta(t - t_0)$ is the Heaviside step function. In both approaches, the amplitude of the radiation field at the exit of the resonant absorber of physical thickness $l$ is

$$a(l, t - t_0) = \frac{1}{2\pi} \int_{-\infty}^{+\infty} A_0(\nu) e^{-i(\omega_S + \nu)(t - t_0) - \alpha(\nu)l} d\nu, \qquad (2)$$

$$A_0(\nu) = \frac{i}{\nu + i\gamma}, \qquad (3)$$

$$\alpha(\nu) = \frac{i\gamma \alpha_B / 2}{\nu + \omega_S - \omega_A + i\gamma_A}, \qquad (4)$$

where $l/c$ is neglected, $A_0(\nu)$ is the Fourier transform of the field amplitude (1), $\alpha(\nu)$ is the function, which describes resonant interaction (absorption and dispersion), $\omega_A$ and $\gamma_A$ are the resonant frequency and coherence decay rate of 14.4 keV transition in the absorber nuclei, and $\alpha_B$ is the Beer's law absorption coefficient for monochromatic radiation tuned in exact resonance. This result is obtained by omitting other non-resonant losses and contribution of radiation fraction with recoil. They can be easily taken into account. Equations (2) - (4) are derived by solving Maxwell-Bloch equations in ref. 33.

In exact resonance ($\omega_S = \omega_A$) and for equal coherence decay rates, $\gamma_A = \gamma$, the integral in (2) is reduced to

$$a(l, t - t_0) = \theta(t - t_0) e^{-(i\omega_S + \gamma)(t - t_0)} J_0\left[2\sqrt{b(t - t_0)}\right], \qquad (5)$$

where $J_0(x)$ is the zero-order Bessel function, $b = T\gamma/2$, and $T = \alpha_B l$ is the optical thickness of the absorber. The field $a(l, t - t_0)$ decays noticeably faster than $\exp[-\gamma(t - t_0)]$ if $T > 1/2$.

To simplify our consideration we use the response function technique[13,14], which helps to express (2) as

$$a(l,t) = \frac{1}{2\pi} \int_{-\infty}^{+\infty} a_0(t-\tau) R(\tau) d\tau , \qquad (6)$$

where

$$R(\tau) = \delta(\tau) - e^{-i(\omega_A + \gamma_A)\tau} \theta(\tau) \sigma_1(\tau) \qquad (7)$$

is the output field from the resonant absorber if the input field is a short pulse, described by the delta function $\delta(t)$, $\sigma_1(\tau) = bJ_1(2\sqrt{b\tau})/\sqrt{b\tau}$, and $J_1(x)$ is the Bessel function of the first order. Here we take $t_0 = 0$ for simplicity.

If the distance between the source and absorber changes as $L(t)$, the phase of the field incident on the absorber also changes as follows $\varphi(t) = 2\pi L(t)/\lambda$, where $\lambda$ is the wavelength of the radiation field. For simplicity we consider instantaneous displacement of the absorber on a half-wavelength at time $t_1$, which changes the field phase by $\pi$. Then, the radiation field seen by the absorber is modified as

$$a_\pi(t - t_0) = \left[\theta(t - t_0) - 2\theta(t - t_1)\right] e^{-(i\omega_S + \gamma)(t - t_0)} , \qquad (8)$$

where it is assumed that $t_1 > t_0$. Substitution of this field into Eq. (6) gives

$$a_\pi(l, t - t_0) = \theta(t - t_0) e^{-(i\omega_S + \gamma)(t - t_0)} \left\{ J_0\left[2\sqrt{b(t - t_0)}\right] - 2\theta(t - t_1) J_0\left[2\sqrt{b(t - t_1)}\right] \right\}, \qquad (9)$$

when $\omega_A = \omega_S$ and $\gamma_A = \gamma$. The plots of the detection probability of the source field $P_0(t) = |a_0(t)|^2$ and the detection probability $P_\pi(t) = |a(l,t)|^2$ of the field transmitted through the absorber experiencing fast displacement at time $t_1$ are shown in Fig.A1. It is clearly seen that initially the transmitted field experiences fast decay due to destructive interference with the field coherently scattered by resonant nuclei. Then, at time $t_1$ this interference becomes constructive producing a radiation flash. Generalization for the case when the absorber displacement takes finite time is given in refs. 13, 14.

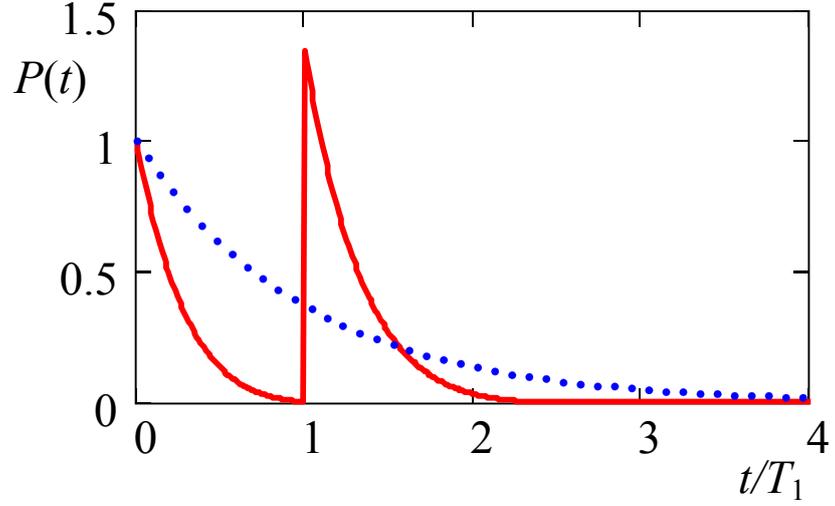

Fig. A1. Dotted blue line shows the exponentially decaying detection probability of the source photon and red solid line shows the detection probability of the photon transmitted through the absorber experiencing fast displacement at time $t_1 = T_1$, where $T_1 = 1/2\gamma$ is a lifetime of the excited nucleus in the source. Optical thickness of the absorber is $T = 5$ and $t_0 = 0$.

Time $t_0$ integrated photon detection probability

$$N_\pi(l,t) = \int_{-\infty}^{+\infty} |a_\pi(l, t - t_0)|^2 dt_0, \qquad (10)$$

describes detector count rate dependence on time. For a single photon wave packet $a_0(t - t_0)$ this integral gives

$$N_0 = \int_{-\infty}^{+\infty} |a_0(t - t_0)|^2 dt_0 = \frac{1}{2\gamma}, \qquad (11)$$

which is just equal to the lifetime $T_1$ of the excited state $^{57}$Fe in the source. This equation means that in time interval $T_1$ one photon is present if one source nucleus was excited.

Calculation of the integral in Eq. (10) gives

$$N_\pi(l,t)/N_0 = n_B + 4\theta(t - t_1)\sigma_0(t - t_1)F_T(t - t_1), \qquad (12)$$

where $n_B = \exp(-T/2)I_0(T/2)$ is a detection probability of photons transmitted through the absorber not experiencing displacements ($n_B$ gives the transmission baseline), $I_0(T/2)$ is the modified Bessel function of zero order; $\sigma_0(t) = J_0\left(2\sqrt{bt}\right)$, and

$$F_T(t-t_1) = e^{-2\gamma(t-t_1)}\sigma_0(t-t_1) - e^{-T/4} + 2\gamma \int_0^{t-t_1} e^{-2\gamma x}\sigma_0(x)dx. \tag{13}$$

This time dependence of the detection probability of photons transmitted through the absorber, $N_\pi(l,t)$, is shown in Fig. A2. Stepwise displacement of the absorber takes place at $t = T_1$. For optical thickness $T = 5$, maximum value of photon counts exceeds the base line $n_B N_0 = 0.27N_0$ by $[4(1 - e^{-T/4})]N_0 = 2.854N_0$, i.e., nearly 10 times.

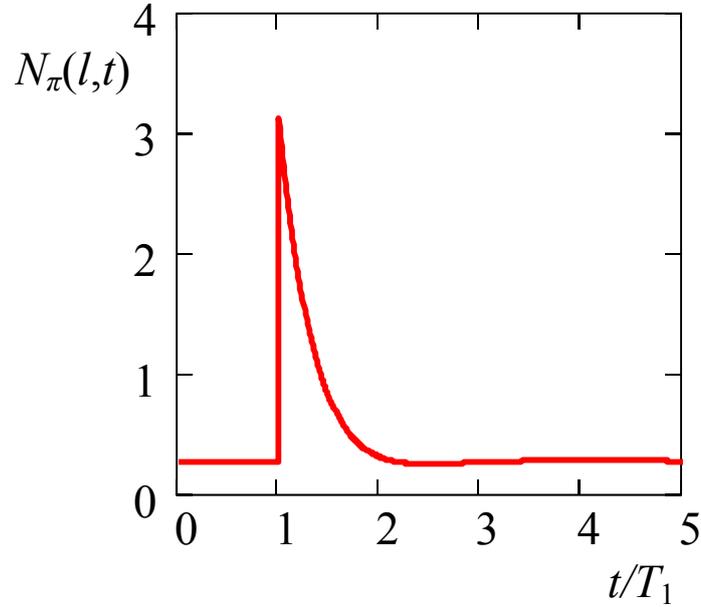

Fig. A2. Time dependence of the detection probability $N_\pi(l,t)$ normalized to $N_0$. Stepwise displacement of the absorber takes place at time $t_1 = T_1$. Optical thickness of the absorber is $T = 5$.

Actually, there are many experimental factors decreasing this large contrast between signal and background. First, the source emits also γ-photons with recoil, which do not interact resonantly with the absorber. Second, we have nonresonant absorption of Mössbauer photons by solid matrix where $^{57}$Fe nuclei are incorporated. Roughly, these factors can be taken into account by expression

$$N(t) = \left[(1-f)N_0 + fN_\pi(l,t)\right]e^{-\beta_{nres}}, \tag{14}$$

where $f$ is Lamb-Mossbauer factor describing recoilless fraction and $\beta$ is responsible for nonresonant absorption.

Another quite important factor is how fast the absorber moves to perform displacement on a half-wavelength. In our simplified consideration, given above, we take stepwise shift of the absorber, which happens instantly. Actually, this is idealized picture. Any piezotransducer is characterized by electrical parameters such as capacitance, inductance, and resistance. Below we use a simplified expression describing time dependence of the absorber displacement $D(t)$ induced by ideal rectangular-voltage pulse, where these parameters are taken into account, i.e.,

$$D(t)/a = \theta(t-t_1)\left[1-\theta(t-t_2)\right]\left[1-e^{-(t-t_1)/T_{r\&d}}\right] + \theta(t-t_2)e^{-(t-t_2)/T_{r\&d}}\left[1-e^{-(t_2-t_1)/T_{r\&d}}\right]. \quad (15)$$

Here, $a = \lambda/\{1 - \exp[-(t_2-t_1)/T_{r\&d}]\}$ is the maximum value of the displacement, which produces $\pi$-shift of the field phase, $t_1$ is time of the beginning of the rectangular-voltage pulse, $t_2$ is time when the voltage starts to drop to zero, and $T_{r\&d}$ is a parameter defining the rates of the voltage rise and drop. Comparison of the rectangular-shaped phase change with realistic time evolution is shown in Fig. A3(a).

To describe realistic case of the phase shifts of the radiation field, which produce $\gamma$-pulses, we use equations[26-28] derived for arbitrary time dependence of the phase, i.e.,

$$N_\pi(l,t)/N_0 = 1 - 2b\int_0^{+\infty} dx\, e^{-2\gamma x}\sigma_1(x)\cos\left[\varphi(t)-\varphi(t-x)\right] + \\ + 2b^2 \int_0^{+\infty} dx \int_0^x dy\, e^{-2\gamma x}\sigma_1(x)\sigma_1(y)\cos\left[\varphi(t-x)-\varphi(t-y)\right] \quad (16)$$

where $\varphi(t) = \pi D(t)/\lambda$ is the time dependent phase. Time evolution of the photon counts induced by these phase shifts are shown in Fig. A3 (b) and compared with the ideal case when phase changes instantly.

It is obvious that the presence of $\gamma$-pulses changes the overall number of counts in the time window $T_S$ of data collection. Therefore, the number of counts collected in some time window $T_E$ by eavesdropper will also change when information is transmitted even if $T_E$ does not precisely equal to $T_S$. Since many factors influence on this change, we do not calculate it. This change can be just measured for a particular message. For example, for our message the number of counts, which are collected within a particular time interval without synchronization of sending and reading, is 391 with message and the number of counts is 358 in the same time interval without message. The difference is 9.2%. The absorber becomes slightly more transparent with message. This difference can be compensated by placing a proper filter after the absorber during the message transmission. In our message we have 14 rectangular-voltage pulses with 28 edges producing 28 $\gamma$-pulses. Thus, each rectangular pulse in the message gives 0.66% contribution to bleaching of the absorber. For a new message, one can calculate the percentage of the absorber "bleaching" easily by counting the number of rectangular-voltage pulses in the message. Then, it is easy to find a proper filter hiding the presence of the information transmission.

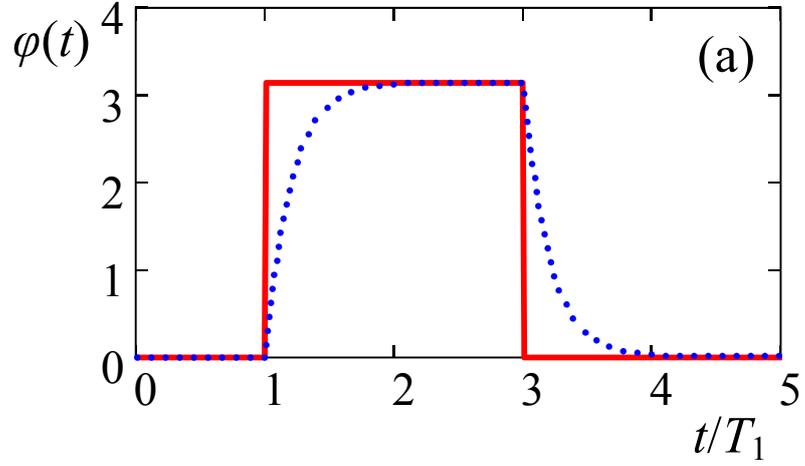

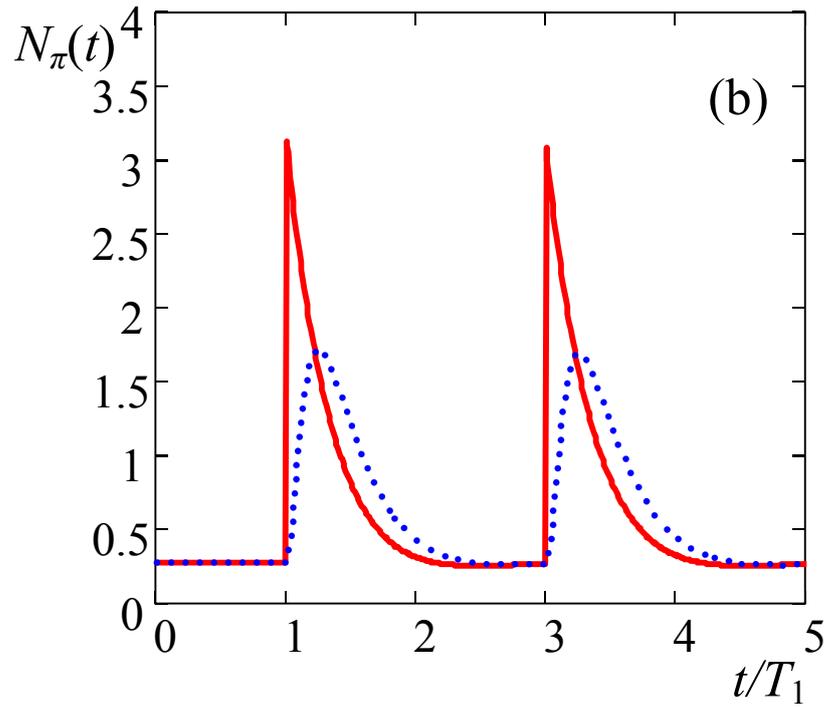

Fig. A3. (a) Time dependence of the phase (in radians) of the radiation field produced by mechanical displacement of the absorber with respect to the source; $t_1 = T_1$ is time when the applied voltage starts to rise and $t_2 = 3T_1$ is time when it starts dropping. Idealized case is shown by red solid line and realistic time dependence is shown by blue dots. For the latter the maximum excursion of the phase is $\pi/\{1 - \exp[-(t_2-t_1)/T_d]\}$. $T_{r\&d} = 0.2T_1$ is the parameter defining the rates of the phase rise and drop. (b) Time dependence of the detection probability (normalized to $N_0$) for ideal case (red solid line) and for realistic case (blue dots). The parameters are the same as in (a).

Summarizing, we conclude that in $\gamma$-protocol the sender produces bleaching of the absorber at very short time intervals for $\gamma$-quanta. Pulses generated in these intervals contain information. Since photon emission is a random process distributed evenly in time, the bleaching of the absorber in a form of $\gamma$-pulses cannot be directly detected. To disclose the information, generation of pulse trains is periodically repeated and detection is synchronized with this period. Collecting many counts synchronously with the sender, recipient can read the message. Synchronization must be very precise. For classical field, for example, CW laser radiation, which is stepwise phase modulated by electrooptical modulator and transmitted through a resonant absorber, one can also generate a series of pulses[32]. However, these pulses are visible and information transmission is not secure.